\documentclass[twocolumn,floatfix,showpacs,amsmat]{revtex4}
\usepackage{graphics,epsfig,graphicx}
\usepackage{color}
\usepackage{makeidx}
\usepackage{latexsym}
\usepackage{calrsfs}
\usepackage{gensymb}

\begin{document}

\title{Quantum interference from remotely trapped ions}

\date{\today}


\author{S. Gerber, D. Rotter, M. Hennrich, R. Blatt$^*$} \affiliation{
Institute for Experimental Physics, University of Innsbruck,
Technikerstr.\ 25, A-6020 Innsbruck, Austria \\
$^*$ Institute for Quantum Optics and Quantum Information of the
Austrian Academy of Sciences, Innsbruck, Austria}

\author{F. Rohde, C. Schuck, M. Almendros, R. Gehr, F. Dubin, J. Eschner}

\affiliation{
Institute of Photonic Sciences (ICFO), Mediterranean Technology Park,\\
Av. del Canal Olimpic s/n, 08860 Castelldefels, Spain }

\pacs{42.50.Ar, 42.50.Ct, 42.50.Dv}

\begin{abstract}
We observe quantum interference of photons emitted by two continuously laser-excited
single ions, independently trapped in distinct vacuum vessels. High contrast two-photon
interference is observed in two experiments with different ion species, Ca$^+$ and
Ba$^+$. Our experimental findings are quantitatively reproduced by Bloch equation calculations. In particular, we show that the coherence of the individual resonance fluorescence light field is determined from the
observed interference.
\end{abstract}
\maketitle


\section{Introduction}

Interfacing static quantum bit registers and quantum communication channels is at the
focus of current research efforts aiming at the realization of quantum networks. For
implementing static qubits, trapped ions are known to be promising candidates as they
provide optimum conditions for quantum information processing: they offer long coherence
times for information storage \cite{Haeffner2005}, are easily manipulated coherently, and
their Coulomb interaction can be utilized to realize quantum logic gates
\cite{CiracZoller1995, 2IonGate2003}. Single photons, on the other hand, are well
suited for transmitting quantum information over long distances
\cite{SinglePhotonQiTransmission}. Hence, ion
traps may constitute the nodes of a quantum network, while communication between remote
nodes may be achieved through photonic channels transmitting quantum states and
entanglement~\cite{Cirac1997}.

Several experiments have recently addressed the realization of an atom-photon qubit
interface and its basic components. Studies were performed with various systems,
including atom-cavity devices \cite{Wilk, Legero, Nussmann2005, Kimble2008,Kimble2007, Keller2004},
atomic ensembles \cite{atom-light-ent, vuletic, storage-retrieval}, and single trapped
atoms or ions \cite{blinov, Grangier, CM_interference, 2photon, CM_entangle}. With these
latter systems, the reported experimental progress comprises: single-photon interference
between two ions in the same trap and continuously excited by a laser \cite{Eichmann1993, Eschner2001}, interference between photons emitted by two atoms in distinct traps under pulsed laser-excitation  \cite{Grangier}, two-photon interference in single-ion fluorescence under continuous excitation \cite{2photon} ,
and finally two-photon interference between two ions confined in remote traps excited by short laser pulses
\cite{CM_interference}. The latter experiments have very recently lead to the demonstration of remote
entanglement between the two independently trapped ions \cite{CM_entangle,
CM_bellstate}.

Entanglement between two remotely trapped atomic systems can be established through
single- or two-photon interference. In the first case, two indistinguishable scattering
paths of a single photon interfere, such that its detection leads to entanglement of the
atoms \cite{1ph-int}. In the latter case, coincident detection of two photons, each being
entangled with one atom \cite{blinov}, projects the two atoms into an entangled state
\cite{FengDuanSimon}. A detailed comparison of the two methods is found in
\cite{Zippilli2008}.

As originally shown by Hong, Ou and Mandel \cite{Mandela}, two indistinguishable photons
impinging simultaneously on two input ports of a beam splitter show photon coalescence,
i.e.\ both photons will leave the beam splitter in the same output mode.
Indistinguishability of the input photons is then marked by a vanishing coincidence rate
between the two output ports, corresponding to the generation of two-photon states.
The contrast of two-photon interference is reduced, however, when the spatial or temporal
overlap of the incident photon wave packets is only partial, such that the photons become
distinguishable \cite{Legero}.

In the experiments reported here we study the conditions to achieve entanglement between distant atoms with two promising systems, Ba$^+$ and Ca$^+$ ions. We observe and characterize quantum interference between photons emitted by two ions of the same species which are independently trapped in different vacuum chambers, approximately one meter apart. Under continuous laser excitation, resonance fluorescence photons are collected in single-mode optical fibers and overlapped at two input ports of a beam splitter. At the output ports, correlations among photon detection events are evaluated. Controlling the polarization at the beam splitter inputs, we observe high-contrast two-photon interference. Recorded correlations allow us to quantify the coherence of the single-ion resonance fluorescence under continuous excitation.

\section{Experimental apparatus}

To realize the experiments, we developed a new assembly which integrates a linear Paul trap and two high-numerical-aperture laser objectives (HALO lenses). The diffraction-limited lenses are mounted inside the vacuum vessels on opposite sides of the trap; their positions are individually controlled by 3-axes piezo-mechanical actuators and allow us to collect up to 8$\%$ of the total fluorescence (4 $\%$ with each lens). The principal components of the experimental set-up are described in the following.

\subsection{HALO lenses}

Each custom-designed objective for collecting the fluorescent light covers a numerical aperture of NA=0.4 with a focal length of 25mm, thus collecting $4\%$ of the total solid angle. The objectives are designed to be diffraction limited over a wide spectral range from ultraviolet to near infrared. The back focal length varies between 13.7mm ($@$400nm) and 15.1mm ($@$870nm), leaving several millimetres of space between the radio frequency electrodes of the ion trap and the front surface of the objective. These consist of four lenses which are all anti-reflection coated for the respective wavelengths (barium: 493nm, 650nm, 1762nm; calcium: 397nm, 422nm, 850-866nm). The calculated transmission of each HALO objective is for calcium: $98.8\%$ (850-866nm), 87.5\% (@729nm), 97.2\% (@422nm), 95.7\% (@397nm); for barium 97\% (@650nm), 98\% (@493nm). The objectives are mounted on vacuum compatible slip-stick piezo translation stages, (Attocube {\it xyz}-positioner, ANPxyz 100). These exhibit a travel of 5mm and 400 nm step size which allows us to maximize the collection efficiency of emitted resonance fluorescence and to address single ions individually when loading the trap with a string of ions.

\begin{figure}[t!] \begin{center}
\includegraphics[width=8.5cm]{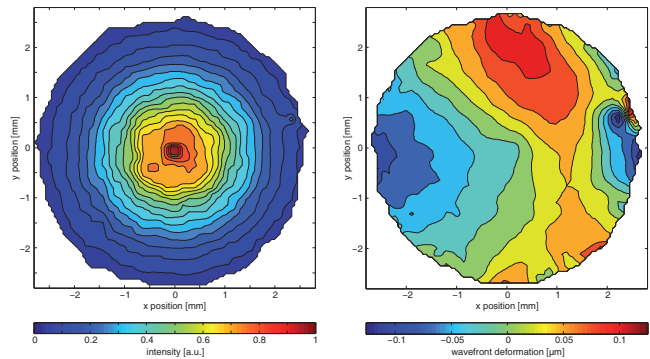}
\caption{ (Color online)
Characterisation of the high-numerical aperture lenses using a wavefront sensor. A 670nm laser is coupled through a fiber. The fiber output is collimated by a HALO lens, and coupled through the setup consisting of the two opposing HALO lenses and the ion trap. The again collimated beam is sent to a wavefront sensor after passing the three HALO lenses. The beam covers 25\% of the exit aperture of the lenses. Left: Gaussian beam profile on the wavefront sensor with a beam waist of 2.3mm. Right: Wavefront aberration of the beam giving an RMS value of 0.049 $\mu$m = 0.073$\lambda$ (@670nm).
\label{HALOwf}}
\end{center}
\end{figure}

 The objectives have been characterised by various methods: For the first batch of objectives, of which one is used in the barium experimental setup,  the manufacturer characterised the wavefront aberrations with a Zygo interferometer to be smaller than $\lambda/10$ at 493nm. Interference experiments proved this precision by obtaining single-ion self-interference with a $72\%$ visibility \cite{Eschner2001}. The second batch of objectives was characterised using a wavefront sensor which showed a root mean square aberration of 0.073$\lambda$ at 670nm for three consecutive HALO lenses, measured over $25\%$ of the aperture as depicted in Fig.~\ref{HALOwf}.

\subsection{Linear trap design}

In the following experiments we use a macroscopic linear trap, identical to the one described in \cite{Gulde_diss}.

\subsection{Experimental setup for barium ions}

\begin{figure}[t!] \begin{center}
\includegraphics[width=8.5cm]{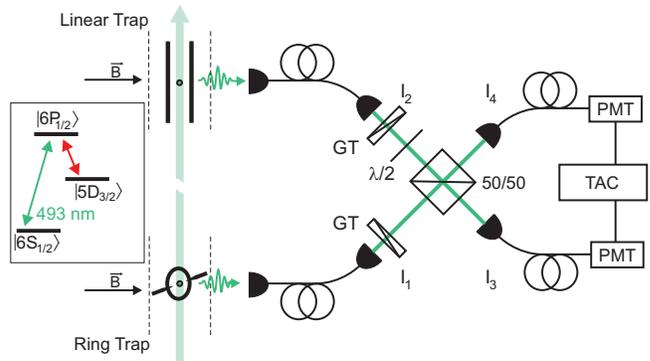} \caption{ (Color online)
Sketch of the barium experimental setup (see text for details). Resonance fluorescence photons emitted by each ion are overlapped on a 50/50 beam splitter (GT stands for Glan-Thompson polarizers). Subsequently, the second-order correlation among photon detections is evaluated. Single photon arrival times are monitored with a Time to Amplitude Converter (TAC) with 100 ps resolution.  The inset shows the relevant electronic
levels of $^{138}$Ba$^{+}$.} \label{setupIBK}
\end{center}
\end{figure}

The schematic experimental setup, located in Innsbruck, and the relevant level scheme of $^{138}$Ba$^+$ are
shown in Fig.~\ref{setupIBK}. One single Ba$^+$ ion is loaded into a ring trap (ion 1), a
second ion into a linear trap (ion 2) using photo-ionization with laser light near 413~nm
\cite{Rotter_diss}. The ions are continuously driven and laser-cooled by the same
narrow-band tunable lasers at 493~nm (green) and 650~nm (red) exciting the
$|\textrm{S}_{1/2}\rangle\rightarrow |\textrm{P}_{1/2}\rangle$ and
$|\textrm{P}_{1/2}\rangle\rightarrow |\textrm{D}_{3/2}\rangle$ transitions, respectively.
After Doppler cooling, the ions are left in a thermal motional state in the Lamb-Dicke
regime \cite{motion}. Laser parameters as well as magnetic fields at the location of the
ions are calibrated by recording excitation spectra \cite{Toschek}.

From each ion, and using the previously described HALO lenses, we collect $4 \%$ of the green resonance fluorescence which is mode-matched into single mode optical fibers. Photons are then guided to the entrance ports of a 50/50
beamsplitter labeled by $I_{1}$ and $I_{2}$ corresponding to ion 1 and ion 2,
respectively (see Fig. \ref{setupIBK}). The two output ports are denoted by $I_{3}$ and
$I_{4}$. Note that the polarization of the two fiber outputs are controlled with
Glan-Thompson polarizers to cancel out polarization fluctuations of the fiber
transmission. The polarization of photons from ion 1 is set horizontal, while the polarization of photons from the second ion is varied. Finally, the beam splitter output ports are
again coupled into single mode fibers to guarantee optimum spatial mode overlap.
Correlations are obtained by subsequently monitoring and correlating the arrival times of
photons at the two PMTs in a Hanbury-Brown and Twiss setup with sub-nanosecond time
resolution. The visibility of the interferometer is measured to be 97~$\%$ by sending
laser light through the input fibers.

\subsection{Experimental setup for calcium ions}

\begin{figure}[t!] \vspace{10mm} \begin{center}
\includegraphics[width=9cm]{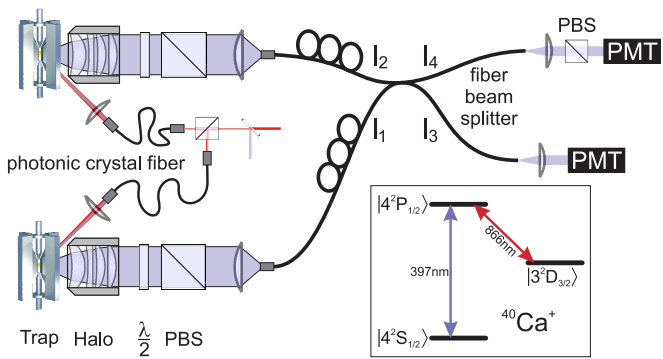} \caption{ (Color online)
Sketch of the calcium experimental setup (see text for details). Fluorescence photons are coupled into the two
inputs of a single mode fiber beam splitter. Stable polarization is then ensured using
fiber polarization controllers (successive loops) and optionally a polarizing beam-splitter (PBS) for control. Second order correlations are computed by recording arrival times at the detectors (PMT) using single-photon counting
electronics. The overall time resolution of the detection setup is 1.5 ns. The inset shows the relevant electronic levels of a
$^{40}$Ca$^{+}$ ion.} \label{setupBCN}
\end{center}
\end{figure}

The schematic of the setup for the $^{40}$Ca$^+$ experiment, located in Barcelona, and the corresponding level
scheme are shown in Fig.~\ref{setupBCN}. The setup consists of two linear Paul traps
mounted in two vacuum vessels separated by about one meter. In each of the two traps a
single $^{40}$Ca$^+$ ion is loaded. For this purpose a novel two-step photo-ionization
method has been developed \cite{carsten} which involves a single-pass frequency doubling
source of 423~nm light and a high-power LED at 389~nm. The ions are continuously excited
and cooled on the $|\textrm{S}_{1/2}\rangle \rightarrow |\textrm{P}_{1/2}\rangle$
electronic transition by a laser at 397~nm. A second laser at 866~nm resonant with
the $|\textrm{P}_{1/2}\rangle \rightarrow |\textrm{D}_{3/2}\rangle$ transition prevents
optical pumping into the metastable $|\textrm{D}_{3/2}\rangle$ state. Both lasers are
locked via transfer cavities to a reference laser, which itself is stabilized to
an atomic -- cesium -- line via Doppler-free absorption spectroscopy \cite{felix}. Excitation
parameters, e.g. laser intensities and frequencies, are set to maximize the count rate
of the fluorescence light while maintaining sufficient laser cooling.

Fluorescence photons from both ions are collected using the previously described HALO
lenses and then coupled into the two input ports of a single-mode fiber beam splitter. A
polarizer before the input coupler projects onto horizontal polarization. Thereafter,
fiber polarization controllers give full and independent control over the polarizations
arriving at the two beam splitter inputs. The use of a fiber beam splitter instead of a
free-space setup ensures optimum spatial overlap of the photon wave-packets
at the beam splitter.

At the outputs of the fiber beam splitter, two photomultipliers with $\eta \approx 0.25$
quantum efficiency detect arriving photons. Their arrival times are recorded with
picosecond resolution (PicoHarp 300 photon counting electronics). The transient time spread of the photomultipliers sets the overall time resolution in this 
experiment to about 1.5~ns.

\section{Theoretical description}

In this part, we present a basic theoretical description to quantify two-photon interference for our experimental configuration. As depicted in Figures \ref{setupIBK} and \ref{setupBCN}, we label the two input ports of the beam splitter as $I_{1}$ and $I_{2}$. The corresponding fields, $\widehat E_{1}$ and $\widehat E_{2}$ are written in an ($\vec e_x$, $\vec e_y$) polarization basis. In the following, fluorescence photons emitted on the $|{\rm P}_{1/2}\rangle$ to $|{\rm
S}_{1/2}\rangle$ electronic transition are expressed in terms of Pauli lowering operators, $\hat\sigma^-_1$ and $\hat\sigma^-_2$, for ion 1 and 2, respectively. These are associated with the creation of a single photon. Without any loss of generality we assume that $\hat\sigma^-_1$ transforms the field into $x$ polarization such that
\begin{eqnarray}
\widehat E_{1}(t)&=&\hat\sigma^-_1(t)~\vec e_x \\ \nonumber
\widehat E_{2}(t)&=&\hat\sigma^-_2(t)(\cos\phi~\vec e_x+\sin\phi~\vec e_y),
\end{eqnarray}
where $\vec e_x$ and $\vec e_y$ denote unit vectors in $x$ and $y$ directions, respectively. In the previous expressions, let us note that the source free part of the input fields has been neglected since noise contributions play a minor role in our experiments \cite{DZ}.

For an ideal 50/50 beam-splitter the transmission ($t$) and reflection ($r$) coefficients
obey $-i\cdot t= r=\frac{1}{\sqrt2}$ \cite{note2}. Therefore, field operators at the
output arms read \newpage
\begin{eqnarray}
\widehat E_3(t)&=&\frac{1}{\sqrt2}\big\{ [\hat\sigma^-_1(t)e^{i\psi}+i\cos\phi \hat\sigma^-_2(t)]\vec
e_x+[i\sin\phi \hat\sigma^-_2(t)]\vec e_y\big\}  \nonumber\\
\widehat E_4(t)&=& \frac{1}{\sqrt2}\big\{ [i\hat\sigma^-_1(t)e^{i\psi}+\cos\phi
\hat\sigma^-_2(t)]\vec e_x+[\sin\phi \hat\sigma^-_2(t)]\vec e_y\big\},\nonumber\\
\end{eqnarray}
where $\psi$ represents a random phase between the fields emitted by the two ions. In fact, our experimental setup does not ensure sub-wavelength mechanical stability which imposes the inclusion of $\psi$.
The overall correlation function can then be expressed as $G^{(2)}_{\rm tot}(t,t+\tau)\propto \sum_{(i,j)=\{x,y\}} \langle\widehat{E}^{\dag}_{3,i}(t)\widehat{E}^{\dag}_{4,j}(t+\tau)\widehat{E}_{4,j}(t+\tau)\widehat{E}_{3,i}(t)\rangle$ where  $\widehat{E}^{\dag}_{(3,4),x}$ corresponds to the part of the field $\widehat{E}^{\dag}_{(3,4)}$ polarized along $x$. Let us now assume that the emission properties of ions 1 and 2 are identical such that after averaging over all possible values of $\psi$ one finds
\begin{eqnarray}
G^{(2)}_{\rm{tot}}(t,t+\tau,\phi)&\propto&\frac{1}{4}[\langle \hat\sigma^+_{1}(t)
\hat\sigma^+_{1}(t+\tau)\hat\sigma^-_{1}(t+\tau)\hat\sigma^-_{1}(t)\rangle
\nonumber\\
&+&\langle \hat\sigma^+_{2}(t) \hat\sigma^+_{2}(t+\tau)
\hat\sigma^-_{2}(t+\tau)\hat\sigma^-_{2}(t)\rangle \nonumber\\
&-&2\cos^2(\phi)\langle\hat\sigma^+_{1}(t)
\hat\sigma^-_{1}(t+\tau)\rangle \nonumber\\
&\cdot&\langle\hat\sigma^+_{2}(t+\tau)\hat\sigma^-_{2}(t)\rangle \\
&+&2\langle \hat\sigma^+_{1}(t+\tau) \hat\sigma^-_{1}(t+\tau) \rangle\langle
\hat\sigma^+_{2}(t) \hat\sigma^-_{2}(t) \rangle \nonumber
].
\end{eqnarray}

The first two terms in Equation (3) represent second order
correlations between photons both emitted by the same ion, ion 1 and 2 for the first and second term respectively. The third term shows the interference between two photons emitted by different ions. It can be rewritten as a product of the individual first-order correlation functions, $G^{(1)}$, as
$-2\cos^2(\phi)~[G^{(1)}_{1}(t,t+\tau)~G^{(1)}_{2}(t,t+\tau)^\ast]$. This term reflects the degree of indistinguishability of photons at the input ports of the beam splitter and consequently vanishes for $\phi=\pi/2$, i.e. for orthogonal polarizations.
The forth term represents the level of random correlations between photons emitted by different ions. This contribution reads as the product of the mean number of photons emitted by each ion, which we denote $\langle n\rangle^2$.
Under the assumption of identical emission properties for ions 1 and 2 we can
set $G^{(2)}_1(t,t+\tau)$=$G^{(2)}_2(t,t+\tau)$=$G^{(2)}(t,t+\tau)$ and $G^{(1)}_1(t,t+\tau)$=$G^{(1)}_2(t,t+\tau)$=$G^{(1)}(t,t+\tau)$ and obtain
\begin{eqnarray}
G^{(2)}_{\rm{tot}}(t,t+\tau,\phi)=\frac{1}{2}[G^{(2)}(t,t+\tau)-\nonumber\\\cos^2(\phi)|G^{(1)}(t,t+\tau)|^2
+\langle
n\rangle^2]. \label{eq4}
\end{eqnarray}
At steady state ($t \rightarrow\infty$), the normalized correlation function is given by $g^{(2)}_{\rm{tot}}(\tau,\phi)=G^{(2)}_{\rm{tot}}(\tau,\phi)/G^{(2)}_{\rm{tot}}(\tau\rightarrow\infty,\phi)$, and it
follows
\begin{equation}
g^{(2)}_{\rm{tot}}(\tau,\phi)=\frac{1}{2}g^{(2)}(\tau)+\frac{1}{2}[1-\cos^2(\phi)
|g^{(1)}(\tau)|^2].
\label{eq4}
\end{equation}

In Eq. (\ref{eq4}), we note that the two-photon interference contribution (last term) has an amplitude which is given by the individual coherence properties of the interfering fields, even for $\phi$=0, i.e. for identical polarization. Indeed, $|g^{(1)}|$ continuously decreases, starting from $|g^{(1)}(0)|$=1, with a time constant equal to the coherence time of the resonance fluorescence. Hence, for $\phi$ set to 0, the coincidence rate vanishes due to both, the discrete character of photon emission by a single-ion ($g^{(2)}(0)$=0, anti-bunching) and the complete photon coalescence at the beam-splitter. On the other hand, for $\tau\neq0$, the individual coherence of interfering fields is revealed by the $g^{(2)}_{\rm{tot}}$-function.

\section{Experimental Results}

\begin{figure}[h!]
\includegraphics[height=6cm]{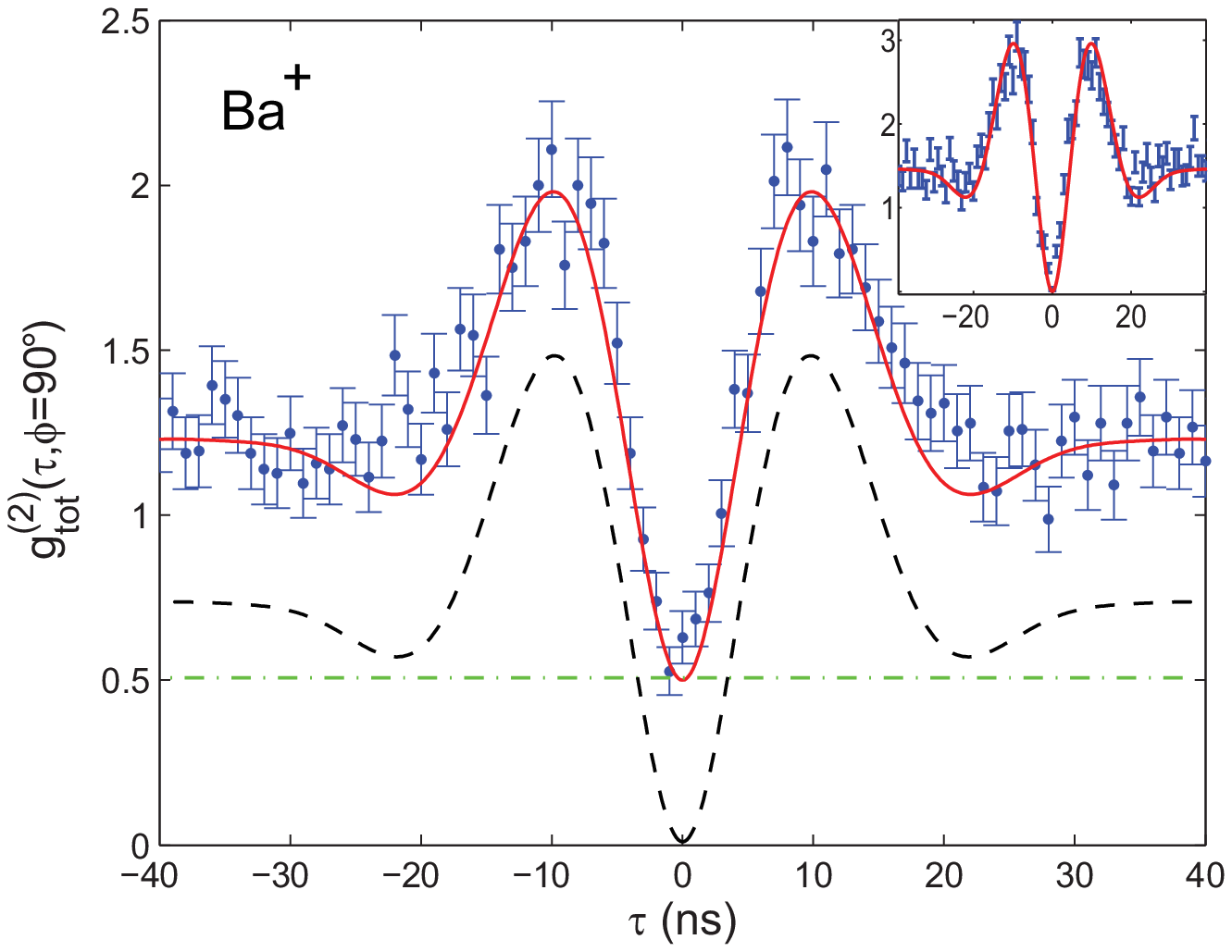}
\includegraphics[width=8cm]{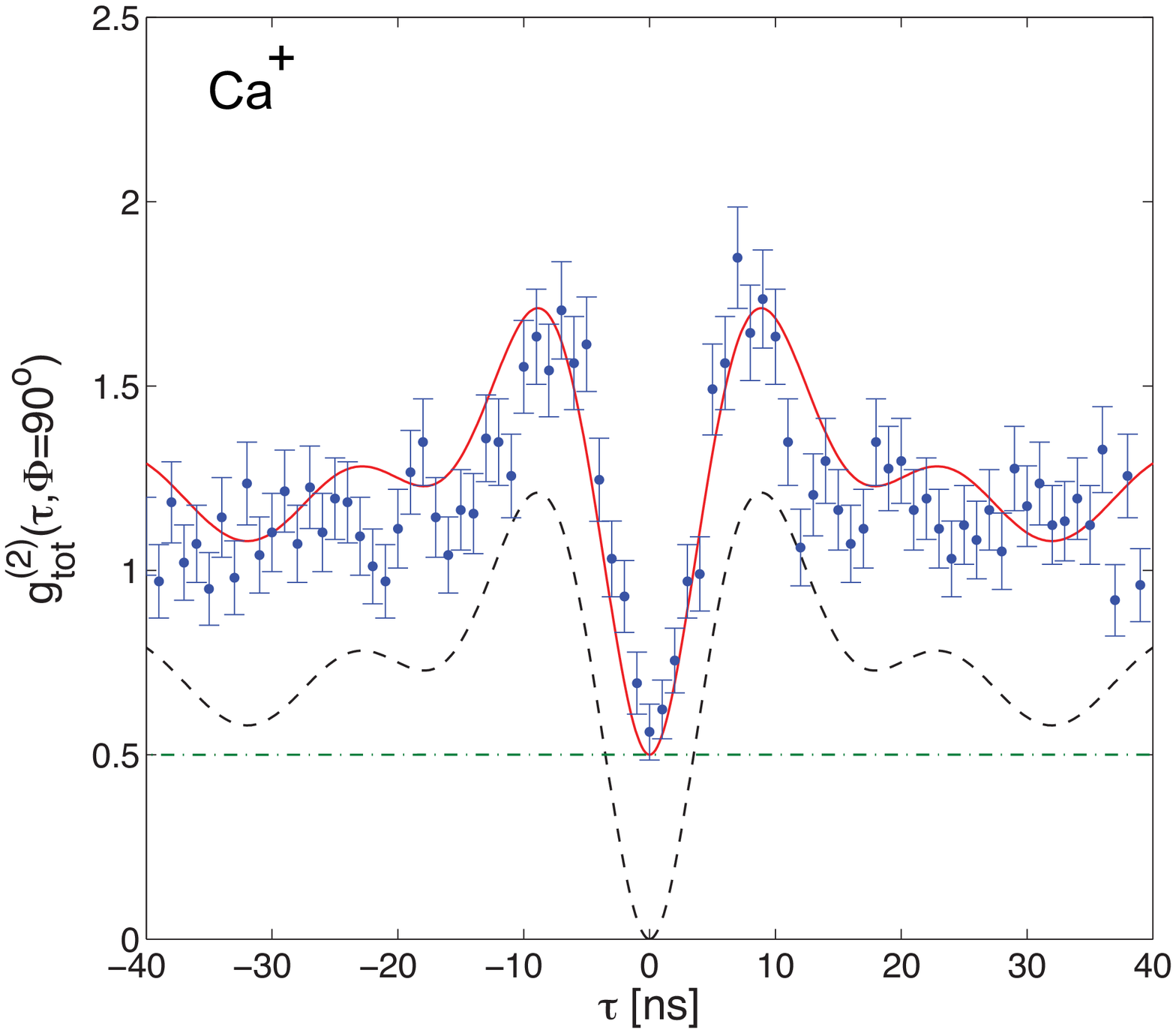}
\caption{(Color online) Normalized second order correlation function for orthogonal polarization between $I_{1}$ and $I_{2}$ ($\phi=90\degree$, i.e the non-interfering case). {\it Top:} Experimental results obtained with the barium experiment. Data points are obtained after an accumulation of 30~minutes. {\it Inset}: Normalized
second order correlation function when one fiber output port is blocked. {\it Bottom:} Experimental results obtained with the calcium experiment. Data are obtained after 2.5 hours of accumulation. In both panels, the result of our theoretical model (Eq. \ref{eq4}) is displayed by the solid line. It is obtained considering our experimental parameters, e.g. laser intensities and detuning, from a full 8-level Bloch equation model. The dashed and dash-dotted lines show the contributions of the first and second term of Eq. (\ref{eq4}), respectively. Experimental data are all presented with 1~ns resolution, without background subtraction, including the variance obtained from shot noise (Poisson statistics at all times $\tau$).} \label{fig3}
\end{figure}

\begin{figure}[h!]\begin{center}
\includegraphics[width=8cm]{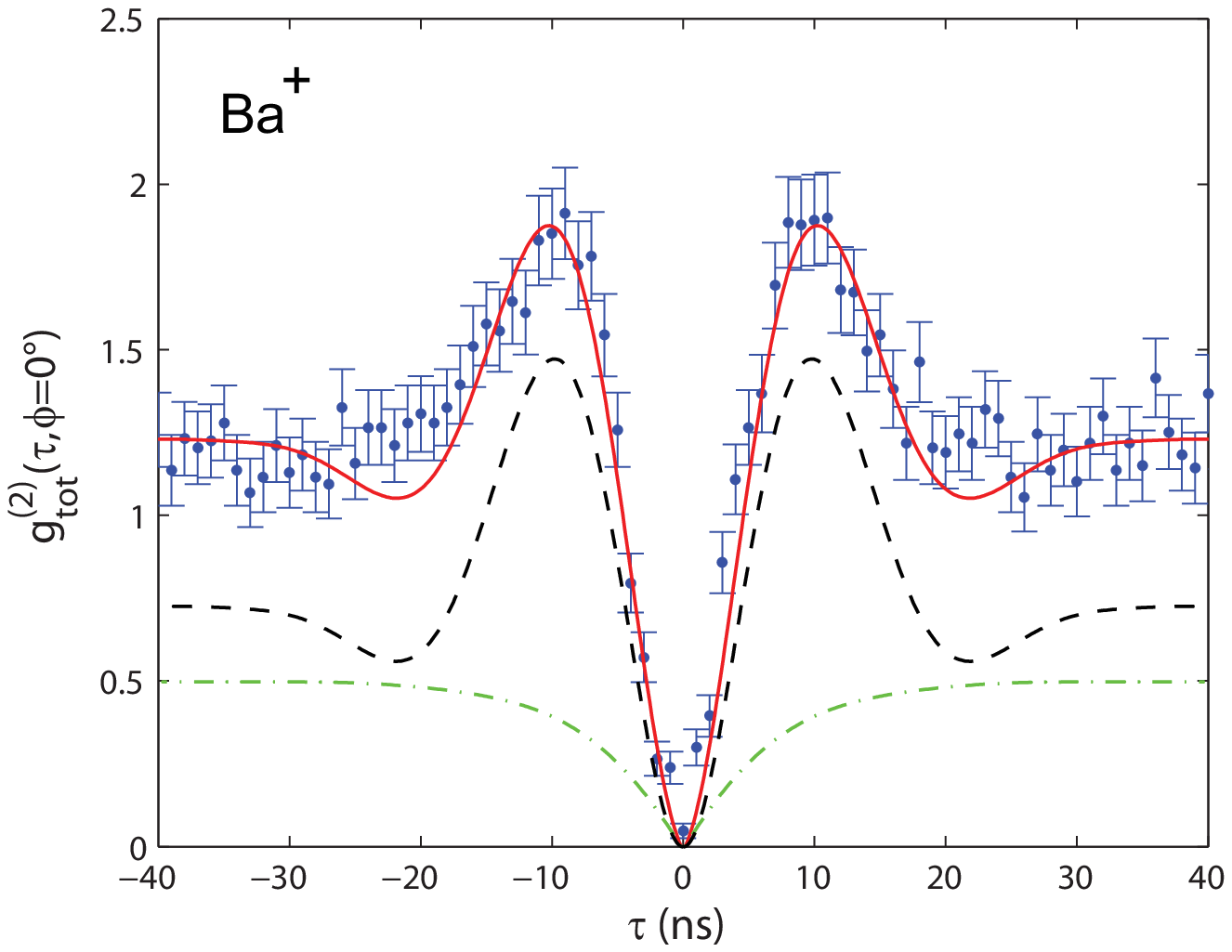}
\includegraphics[width=8cm]{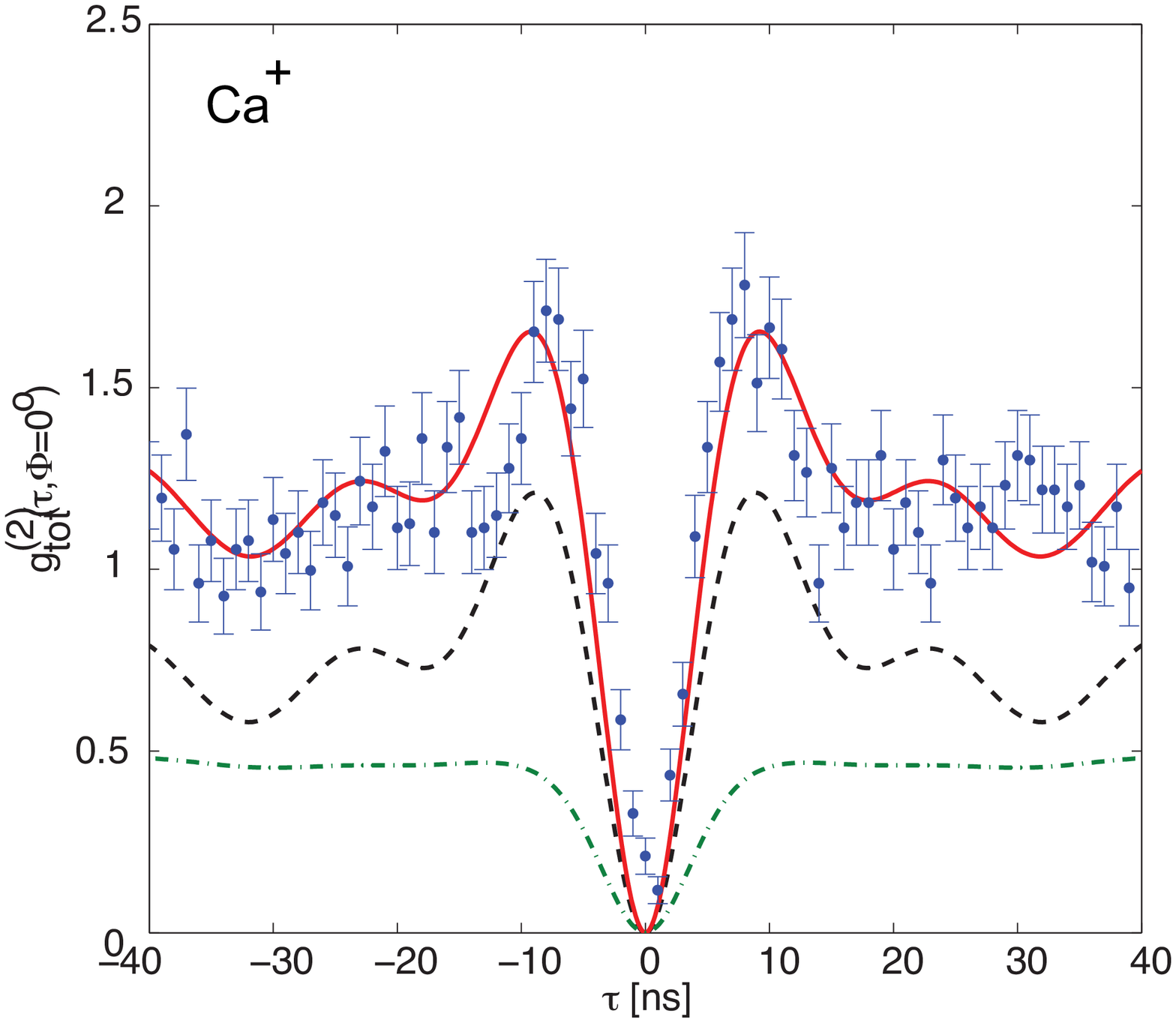} \caption{ (Color online)
Normalized second order correlation function for indistinguishable photons impinging at the input ports of the beam-splitter ($\phi=0\degree$). The top panel presents the results of the barium experiment while the bottom panel shows the experimental results obtained with the calcium experiment. As in Figure \ref{fig3}, the result of our theoretical model (Eq. \ref{eq4}) is displayed by the solid line, the dashed and dash-dotted lines showing the contributions of the first and second term of Eq. (\ref{eq4}), respectively. Experimental data are again presented with 1~ns resolution, without background subtraction, including the variance obtained from shot noise (Poisson statistics). The two-photon interference contrast deduced from these measurements reaches 89(2) and 80(10) $\%$ for the barium and calcium experiment, respectively.} \label{fig4}
\end{center}
\end{figure}

Figure \ref{fig3} presents the second order correlation function, $g^{(2)}_{\rm{tot}}$, when photons emitted by the two ions are made fully distinguishable by setting $\phi=90\degree$. The top (bottom) panel of Fig. \ref{fig3} shows the results obtained with the barium (calcium) experiment. In both cases, the coincidence rate is such that $g^{(2)}_{\rm{tot}}(0,90$\degree$)\approx$ 0.5, as expected from equation (\ref{eq4}). Indeed, for input photons with orthogonal polarization, the last term in Equation (\ref{eq4}) reduces to (1/2) independent on $\tau$. This reflects that among all possible correlations, one half occurs between photons emitted by different ions. These are here rendered distinguishable by their orthogonal polarization such that interference is absent. In Figure \ref{fig3}, the result of our theoretical model (Eq. (\ref{eq4})) is also presented. It is obtained solving the Bloch equations for our experimental parameters, including eight relevant Zeeman electronic levels for the ion internal states while neglecting motional degrees of freedom \cite{Toschek}. There is a good agreement for both measurements. We also present in the inset of Figure \ref{fig3} the normalized second-order correlation function of a single-ion, $g^{(2)}$. It is obtained by blocking one fiber entrance and it shows almost ideal anti-bunching ($g^{(2)}(0)$=0.03 with no background subtraction) and an optical nutation, $g^{(2)}(13 ~ns)$=2.9.

Figure \ref{fig4} shows experimental results obtained for $\phi$=0 such that photons impinging at the two input ports of the beam-splitter are indistinguishable. As expected, the normalized number of coincidences strongly drops around $\tau$=0 which signals two-photon interference \cite{2photon}. From the measured values of $g^{(2)}_{\rm{tot}}(0,90\degree)$ and $g^{(2)}_{\rm{tot}}(0,0\degree)$ we deduce that the two-photon interference contrast reaches 89(2) $\%$ and 80(10) $\%$ for the barium and calcium experiments, respectively. In the latter case, the contrast is limited by the time resolution of the photo-detectors. Contrasts are furthermore derived from raw experimental data, without correcting for accidental correlations induced by stray light photons or dark-counts of the photo-detectors.

As previously mentioned, individual coherence properties of interfering fields modify the overall correlation function for $\phi$=0. First, for $\tau$=0, incident photons are detected simultaneously such that these are fully indistinguishable and therefore the amplitude of two-photon interference is maximal. By contrast, for $\tau\neq$0 the degree of indistinguishability of the input photons is characterized by the temporal overlap of their wave-packets. The length of these wave-packets in fact corresponds to the photon coherence time  which then governs the contrast of the interference \cite{Legero}. This property is reflected by the last term of Eq. (\ref{eq4}) which is also represented by a dash-dotted line in Fig. (\ref{fig4}). It continuously increases before saturating to 1/2 for $|\tau|\approx$ 10 ns. This reveals that the coherence time of the resonance fluorescence does not exceed 20 ns in our experiments. This indicates that incoherent scattering is dominant, far longer coherence times would otherwise be observed. As a consequence, let us note that the amplitude of the optical nutation is reduced by $\approx 5 \%$ compared to the non-interfering case ($\phi=90 \degree$).

\begin{figure}
\begin{center}
\includegraphics[width=8cm]{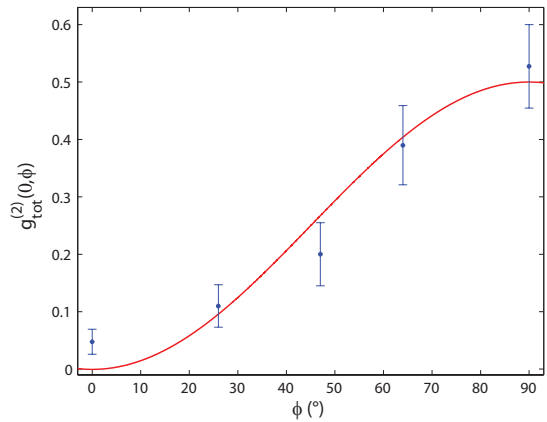} \caption{ (Color online)
Number of coincidences  $g^{(2)}_{\rm{tot}}(0,\phi)$ for varying polarization angles, $\phi\in\{0\degree,~26\degree
,~47\degree,~64\degree,~90\degree\}$. The solid line shows the
$\frac{1}{2}\sin^2(\phi)$ dependence of the coincidence rate at $\tau=0$
as given by Eq. (\ref{eq4}). Experimental results are obtained with the barium experiment.
} \label{fig5}
\end{center}
\end{figure}

Finally, we present in Figure \ref{fig5} the number of coincidence counts as a function of the polarization angle, $\phi$. Thereby the degree of distinguishability of interfering photons, and thus the amplitude of the two-photon interference, is smoothly tuned. From Eq. (\ref{eq4}), one expects a $\frac{1}{2}[1-\cos^2\phi]= \frac{1}{2}\sin^2\phi$ dependence. The latter is represented by a solid line in Fig. \ref{fig5}, in good agreement with our experimental findings.

\section{Summary}

In summary, we have observed high contrast interference between resonance fluorescence photons emitted by remotely trapped ions. Large photon detection efficiency is reached by means of high numerical aperture HALO-lenses placed inside the vacuum vessels. These allow us to reach detection count rates in a single-mode as high as 25-30 and 10-15 kcps/s for the barium and calcium experiment respectively. The discripancy arises from different fiber coupling efficiency and photo-detectors quantum efficiency at 493 and 397 nm. Furthermore, our experiments are quantitatively reproduced by model calculation and allow us to observe the coherence of resonance fluorescence photons by means of two-photon interference under continuous laser excitation.

\end{document}